\documentclass[12pt,reqno]{article}
\usepackage{amsmath}
\usepackage{graphicx}
\usepackage{color}
\usepackage{varioref}

\begin{document}

\title
\centerline{\bf\large{The NNLO QCD analysis of gluon distribution}}\\
\centerline{\bf\large{function at small-$x$}}
\vspace{9pt}

\centerline{{\bf Mayuri Devee}$^{\rm *1}$, {\bf J. K. Sarma}$^{\rm *2}$}
\centerline{\small{$^{\rm 1}$}{Department of Physics, University of Science and Technology Meghalaya, Ri-Bhoi, Meghalaya 793101, India}} 
\centerline{\small{$^{\rm 2}$}{HEP Laboratory, Department of Physics, Tezpur University, Tezpur, Assam 784 028, India}} 
\centerline{$^{\rm *}${\it deveemayuri@gmail.com}}

\renewcommand{\baselinestretch}{1.50}\normalsize

\begin{abstract}
\leftskip0.1mm
\rightskip0.1mm
  A next-to-next-to-leading order (NNLO) QCD calculation of gluon distribution function at small-$x$ is presented. The gluon distribution function is explored analytically in the DGLAP approach by a Taylor expansion at small $x$ as two first order partial differential equations in two variables : Bjorken $x$ and $t$ $(t = \ln(\frac{Q^2}{\Lambda^2})$. We have solved the system of equations at LO, NLO and NNLO respectively by the Lagrange’s method. The resulting analytical expressions are compared with the available global PDF fits as well as with the results of the BDM model. We have further performed a $\chi^2$ test to check the compatibility of our predictions and observed that our results can be consistently described in the context of perturbative QCD. A comparative analysis of the obtained results at LO, NLO and NNLO reveals that the NNLO approximation has a significant contribution to the gluon distribution function particularly in the small-$x$ region.



\end{abstract}

\section{Introduction}

\ The gluon distribution function is one of the extremely indispensable physical observables that controls the physics at high energy or small-$x$ in deep inelastic scattering (DIS), where $x$ is the Bjorken variable. The precise knowledge of gluon distribution functions at small-$x$ provides important insight in the estimation of backgrounds as well as the exploration of new physics at the Large Hadron Collider.
The gluon momentum distribution in the proton can be determined from the quark distributions together with the evolution equations. More direct approach to determine the gluon distribution is based on the reconstruction of the kinematics of the interacting partons from the measurement of the hadronic final state in gluon induced processes. 
The conventional and the fundamental theoretical frameworks employed to study the scale dependence of the parton (quark or gluon) distribution functions (PDFs) and eventually the DIS structure functions are the linear Dokshitzer-Gribov-Lipatov-Altarelli-Parisi (DGLAP) [1-4] evolution equations. It is the basic tool for all of the phenomenological perspectives used to interpret the hadron interactions at short distances. The DGLAP equation sums all leading Feynman diagrams that give rise to the logarithmically enhanced $\ln(Q^2)$ contributions to the cross section in order to neglect any kind of higher twist corrections. The associated perturbative resummation is organized in powers $\alpha_s^n\ln(Q^2)^n$. In consonance with the traditional collinear factorization approach the hadronic observables can be expressed as the convolution of the PDFs with partonic hard-scattering coefficients. 
One can calculate the PDFs for any value of $Q^2$ making use of the DGLAP equations considering that an initial condition for the PDFs is indeed available at a given initial scale $Q_0^2$ and then evolving to higher $Q^2$. The initial distributions are at present have to be computed from experiment presuming an input form in $x$ which complies with the QCD sum rules. This strategy is adopted in the global analyses of PDFs [5, 6]. As an alternative, one may produce the parton distributions dynamically originating from an input distribution for the valence quarks and a valence-like input for the sea quarks and gluons [7].

\ The proton structure function measured by the H1 and ZEUS collaboration at HERA [8-10] over a wide kinematic region makes it possible to know about the gluon distribution in the previously unexplored region of $x$ and $Q^2$. The fast growth of the proton structure function at small-$x$ observed at HERA brings about much attention because perturbative QCD in conjunction with the DGLAP equation [1-4] attributes this sharp growth to a similar rise of gluon density towards small-$x$.
The gluon distribution contributes crucially to the evolution of the parton distribution and governs the structure function $F_2(x,Q^2)$ through the evolution $g\to{q\bar{q}}$ in the small-$x$ region.
The solutions of the unpolarized DGLAP equation for the QCD evolution of structure functions have been discussed considerably over the past years. The standard and the most extensively used procedure of studying the hadron structure functions is via the numerical solution of these equations [11-18], with excellent agreement with the DIS data over a wide kinematic region in $x$ and $Q^2$.
However, apart from the numerical solution, it is very important to explore the possibility of obtaining analytical solutions of DGLAP equations at least in the restricted domain of small-$x$ and many approximated analytical solutions of the DGLAP evolution equations suitable at small-$x$, have been reported in recent years [19-23] with considerable phenomenological success. 

\ Recently, in Refs. [24, 25] the present authors reported the analytical solutions of the singlet structure function upto NNLO in the DGLAP approach at small-$x$. Here, the $Q^2$ and $x$ evolutions of deuteron structure function are also calculated from the predicted solution of singlet structure function upto NNLO with reasonable phenomenological success. In the present paper, we extend the work to derive an explicit expression for the gluon distribution function at LO, NLO and NNLO by solving the corresponding DGLAP evolution equation analytically.
In such an approach, we use a Taylor series expansion valid at small-$x$ and reframes the DGLAP equations as partial differential equations in the variables $x$ and $Q^2$. The resulting equations at LO, NLO and NNLO are then solved by the Lagrange's auxiliary method to obtain the $Q^2$ and $x$-evolutions of the gluon distribution function. Moreover, we investigate the impact of the NNLO contributions on the evolution of the gluon distribution function.
The inclusion of the NNLO contributions considerably reduces the theoretical uncertainty of determinations of the quark and gluon densities from DIS structure functions.
The obtained results can be described within the framework of perturbative QCD. We use the published values of the gluon distributions from the MRST2004NNLO [5], MSTW2008NNLO [6], GRV1998NLO [7] and JR09NNLO [26] global PDF analyses to illustrate the method and check the compatibility of our predicted gluon distributions. We also make a comparative analysis of our predictions with the results of the Block-Durand-McKay (BDM) model [27]. 

\ The plan of the paper is as follows. In section 2 we briefly describe the formalism for the solution of DGLAP evolution equation at LO, NLO and NNLO to study the $Q^2$ and $x$ evolutions of gluon distribution function. We present the results and discussions of our predictions in section 3 and summarize in section 4.


\section{Formalism}
\subsection{General framework}

\ The DGLAP evolution equation for gluon distribution function in the standard form is given by [28]
\begin{equation}\frac{\partial{g}(x,Q^2)}{{\partial}\ln{{Q}^2}}=\int_{x}^{1} \frac{d\omega}{\omega}\Big(P_{gq}(\omega)q_S(x/{\omega},Q^2)+P_{gg}(\omega)g(x/{\omega},Q^2)\Big),\end{equation}
where the splitting function $P_{gq}$ is defined as 
\begin{eqnarray}
P_{gq}(x,Q^2) & = & \frac{{\alpha_s}{(Q^2)}}{2\pi}P_{gq}^{(0)}(x)+\Big(\frac{{\alpha_{s}}{(Q^{2})}}{2\pi}\Big)^2P_{gq}^{(1)}(x)+\Big(\frac{{\alpha_{s}}{(Q^{2})}}{2\pi}\Big)^3P_{gq}^{(2)}(x)\nonumber\\
&& +\: \mathcal{O}P_{gq}^{(3)}(x).
\end{eqnarray}
where $P_{gq}^{(0)}(x)$, $P_{gq}^{(1)}(x)$ and $P_{gq}^{(2)}(x)$ are LO, NLO and NNLO quark-gluon splitting functions respectively.
The gluon-gluon splitting function $P_{gg}$ can be defined in a similar fashion. Here $q_S$ is the singlet quark density and $g$ is the gluon density. The representation $G(x,Q^2)=xg(x,Q^2)$ is used here.

\ The running coupling constant ${\alpha_{S}}{(Q^{2})}$ has the form [14, 32]
\begin{equation}\frac{\alpha_s(Q^2)}{2\pi}=\frac{2}{\beta_0{\ln(Q^2/\Lambda^2)}},\end{equation}
\begin{equation}\frac{\alpha_s(Q^2)}{2\pi}=\frac{2}{\beta_0{\ln(Q^2/\Lambda^2)}}\Big[1-\frac{\beta_1}{\beta_0^2}\frac{\ln\big({\ln(Q^2/\Lambda^2)}\big)}{\ln(Q^2/\Lambda^2)}\Big],\end{equation}
\begin{eqnarray}\frac{\alpha_s(Q^2)}{2\pi}&=&\frac{2}{\beta_0{\ln(Q^2/\Lambda^2)}}\Big[1-\frac{\beta_1}{\beta_0^2}\frac{\ln\big({\ln(Q^2/\Lambda^2)}\big)}{\ln(Q^2/\Lambda^2)}+\frac{1}{\beta_0^3\ln(Q^2/\Lambda^2)}\nonumber\\
&& \: \times\Big\{\frac{\beta_1^2}{\beta_0}(\ln^2\big({\ln(Q^2/\Lambda^2)}\big)-\ln\big({\ln(Q^2/\Lambda^2)}\big)-1)+\beta_2\Big\}\Big]\end{eqnarray}
at LO, NLO and NNLO respectively. Here
\begin{equation}{{\beta_{0}}={{\dfrac{11}{3}}{N_{c}}}-{{\dfrac{4}{3}}{T_{f}}}={11-{{\dfrac{2}{3}}{N_{f}}}}},\nonumber \end{equation}
\begin{equation}{{\beta_{1}}={{{\dfrac{34}{3}}{N_{c}^{2}}}-{{\dfrac{10}{3}}{N_{c}N_{f}}-{2C_{F}{N_{f}}}}={102-{{{\dfrac{38}{3}}{N_{f}}}}}}},\nonumber \end{equation}
\begin{eqnarray}{\beta_{2}}&=&{{{\dfrac{2857}{54}}{N_{c}^{3}}}+{2{C_{F}^{2}}{T_{f}}}-{{\dfrac{205}{9}}{{C_{F}}{N_{c}}{T_{f}}}}-\frac{1415}{27}N_c^2T_f+{{\dfrac{44}{9}}{{C_{F}}{T_{f}^{2}}}}+{{\dfrac{158}{27}}{{N_{c}}{T_{f}^{2}}}}}\nonumber\\
&=&{\dfrac{2857}{2}}-{{\dfrac{6673}{18}}{N_{f}}}+{{\dfrac{325}{54}}{N_{f}^{2}}}\nonumber
\end{eqnarray}
are the one-loop, two-loop and three-loop corrections to the QCD $\beta$-function and $N_f$ being the number of quark flavours. Here we use $N_{f}=4$, $N_{c}=3$. The Casimir operators of the color $SU(3)$ are defined as ${{C_{F}}={\dfrac{{N_{c}^{2}}-1}{2{N_{c}}}}}$ $={\dfrac{4}{3}}$ and ${T_{f}}={{\dfrac{1}{2}}{N_{f}}}$.

\subsection{Analysis of gluon distribution function at LO:}

\ Substituting the explicit form of the LO splitting functions [7, 21] in Eq.(1) and simplifying, the LO DGLAP evolution equation for the gluon distribution function can be written as
\begin{equation}\frac{{\partial}G(x,t)}{{\partial}t}=\frac{\alpha_s(t)}{2\pi}\Big[6\Big(\frac{11}{12}-\frac{N_f}{18}+\ln(1-x)\Big)G(x,t)+6I_1^g(x,t)\Big],\end{equation}
where $G(x,Q^2)=xg(x,Q^2)$ is the gluon distribution function. The integral function $I_1^g(x,t)$ is defined as
\begin{eqnarray}
I_1^g(x,t)&=&\int_x^1d\omega\bigg[\frac{\omega{G\big(\frac{x}{\omega},t\big)}-G(x,t)}{1-\omega}+\Big(\omega(1-\omega)+\frac{1-\omega}{\omega}\Big)G\Big(\frac{x}{\omega},t\Big)\nonumber\\
&& +\: \frac{2}{9}\bigg(\frac{1+(1-\omega)^2}{\omega}\bigg)F_2^S(\frac{x}{\omega}t).\bigg]
\end{eqnarray}
Here the variable $t$ is used where $t=\ln(Q^2/\Lambda^2)$.
Now to simplify and reduce the integro-differential equation to a partial differential equation we introduce a variable $u=1-\omega$ so that the argument $x/{\omega}$ can be expressed as
\begin{equation}{\frac{x}{\omega}={\frac{x}{1-u}}=x+\frac{xu}{1-u}}.\end{equation}
Since $x<\omega<1$, so we have $0<u<1-{x}$. This implies that the above series is convergent for $|u|<1$.
Now using Eq. (8), we can expand ${G(x/\omega,t)}$ by Taylor expansion series as
\begin{equation}{G\Big(\frac{x}{\omega},t\Big)=G(x,t)+\frac{xu}{1-u}\frac{\partial{G(x,t)}}{\partial x}}.\end{equation}
As we consider the small-$x$ ($x<0.1$) domain in our analysis, therefore the terms containing $x^2$ and higher powers of $x$ are neglected in Eq. (9).

Similarly, $F_2^S(\frac{x}{\omega},t)$ can be approximated as
\begin{equation}{F_2^S\Big(\frac{x}{\omega},t\Big)=F_2^S(x,t)+\frac{xu}{1-u}\frac{\partial{F_2^S(x,t)}}{\partial x}}.\end{equation}
Substituting these values of ${G(\frac{x}{\omega},t)}$ and $F_2^S(\frac{x}{\omega},t)$ in Eq.(7) and carrying out the integrations in $u$ we get from Eq.(6)
\begin{eqnarray}\frac{\partial{G(x,t)}}{\partial t}&=&\frac{6A_f}{t}\Big[A_{1}^{g}(x) G(x,t)+ A_{2}^{g}(x)\frac{\partial{G(x,t)}}{\partial{x}}+A_3^{g}(x)F_2^S(x,t)\nonumber\\
&& +\: A_{4}^{g}(x)\frac{\partial{F_2^S(x,t)}}{\partial{x}}\Big],\end{eqnarray}
where $A_i^{g}(x)$ ($i$=1,2,3,4) are functions of $x$ (see Appendix A). {\large $\frac{A_f}{t}=$}{\large$\frac{\alpha_s(t)}{2\pi}$} where $A_f=${\large$\frac{2}{\beta_0}$} and $\beta_0$ is the one-loop correction to the QCD beta function.
Eq.(11) is a partial differential equation for gluon distribution function with respect to the variables $x$ and $t$. Thus using a Taylor expansion valid at small-$x$ we reframe the DGLAP equation for gluon distribution, which is an integro-differential equation, as partial differential equation in two variables $x$ and $t$ or $Q^2$.

\ The gluon parton densities cannot be measured directly through experiments. Therefore the direct relations between $F_2(x,Q^2)$ and the $G(x,Q^2)$ are extremely important because using those relations the values of $G(x,Q^2)$ can be extracted using the data on $F_2(x,Q^2)$. As the gluon distribution is coupled to the singlet structure function, so a relation between gluon distribution function and singlet structure function has to be assumed in order to obtain an analytical solution of the DGLAP evolution equation for gluon distribution function. A plausible way of realizing this is through the commonly used relation ${G(x,t)=K(x)F_2^S(x,t)}$ [15, 16, 19], which gives the possibility to extract the gluon distribution function directly from
the experimental data. Here $K(x)$ is a parameter to be determined from phenomenological analysis.
The evolution equations of gluon parton densities and singlet structure functions are in the same forms of derivative with respect to $Q^2$. Moreover the input singlet and gluon parameterizations, taken from global analysis of PDFs, in particular from the GRV1998, MRST2001, MSTW2008 parton sets, to incorporate different high precision data, are also functions of $x$ at fixed $Q^2$ [33]. So the relation between singlet structure function and gluon parton densities will come out in terms of $x$ at fixed-$Q^2$. Accordingly the above assumption is justifiable. In this analysis we take the function $K(x)$ as an arbitrary parameter $K$ and obtain the best fit results by choosing an appropriate value of $K$ for the satisfactory description of each experiment for a particular range of $x$ and $Q^2$.
From this relation we get $F_2^S(x,t)=K_1(x)G(x,t),$ where $K_1(x)=1/K(x)$.
Using this relation Eq. (11) takes the form
\begin{equation}{-t\frac{\partial{G(x,t)}}{\partial{t}}+L_1^g(x) \frac{\partial{G(x,t)}}{\partial{x}}+M_1^g(x)G(x,t)=0},\end{equation}
with \begin{equation}{L_1^g(x)=6A_f\Big[A_2^g(x)+K_1(x)A_4^g(x)\Big]},\end{equation}
\begin{equation}{M_1^G(x)=\frac{12}{\beta_0}\Big[A_1^g(x)+K_1(x)A_3^g(x)+\frac{\partial{K_1(x)}}{\partial{x}}A_4^g(x)\Big]},\end{equation}
Now the general solution of the equation (12) is \begin{equation}{F(U,V)=0},\end{equation}
where ${F(U,V)}$ is an arbitrary function of $U$ and $V$. Here, ${U(x, t,G(x,t)) = k_1}$ and ${V(x, t, G(x,t))=k_2}$ are two independent solutions of the Lagrange's equation
\begin{equation}{\frac{\partial{x}}{{L_1^g}(x)}=\frac{\partial{t}}{-t}=\frac{\partial{G(x,t)}}{-{M_1^g}(x)G(x,t)}}.\end{equation}
Solving Eq. (16) we obtain
\begin{equation}{ U(x, t, G(x,t)) = t\cdot{\exp}\Big[\int\frac{1}{{L_1^g}(x)}dx\Big]}\end{equation} and
\begin{equation}{ V(x, t, G(x,t)) = G(x,t)\cdot{\exp}\Big[\int\frac{{M_1^g}(x)}{{L_1^g}(x)}dx\Big]}.\end{equation}

\ Thus  we see that it has no unique solution. In this approach we attempt to extract a particular solution that obeys some physical constraints on the structure function. The simplest possibility to get a solution is that a linear combination of $U$ and $V$ should obey the Eq. (15) so that
\begin{equation}{\alpha\cdot{U}+ \beta\cdot{V}=0},\end{equation}
where $\alpha$ and $\beta$ are arbitrary constants to be determined from the boundary conditions on $F_2^{S}$. Putting the values of $U$ and $V$ from Eq.(17) and Eq.(18) respectively in Eq.(19) we get
\begin{equation}{\alpha{t}\cdot{\exp}\Big[\int\frac{1}{{L_1^g}(x)}dx\Big]+\beta{G(x,t)}\cdot{\exp}\Big[\int\frac{{M_1^g}(x)}{{L_1^g}(x)}dx\Big]=0},\end{equation}
which implies,  \begin{equation} {G(x,t)=-\gamma{t}\cdot{\exp}\Big[\int\big(\frac{1}{{L_1^g}(x)}- \frac{{M_1^g}(x)}{{L_1^g}(x)}\big)dx\Big]},\end{equation}
where $\gamma=${\large $\frac{\alpha}{\beta}$} is another constant.

Now defining the initial gluon distributions as
\begin{equation}{G(x,t_0)=-\gamma{t_0}\cdot{\exp}\Big[\int\big(\frac{1}{{L_1^g}(x)}- \frac{{M_1^g}(x)}{{L_1^g}(x)}\big)dx\Big]}\end{equation}
and
\begin{equation}{G(x_0,t)=-\gamma{t}\cdot{\exp}\Big[\int\big(\frac{1}{{L_1^g}(x)}- \frac{{M_1^g}(x)}{{L_1^g}(x)}\big)dx\Big]_{x=x_0}}\end{equation}
at any lower value ${t=t_0}$ and at any higher value ${x=x_0}$ respectively,
Eq. (21) leads us to
\begin{equation} {G(x,t)=G(x,t_0)\Big(\frac{t}{t_0}\Big)}\end{equation}
and
\begin{equation} {G(x,t)=G(x_0,t){\exp}\Big[\int_{x_0}^{x}\big(\frac{1}{{L_1^g}(x)}- \frac{{M_1^g}(x)}{{L_1^g}(x)}\big)dx\Big].}\end{equation}
Thus Eq. (24) gives the $Q^2$-evolution for gluon distribution function at LO at a particular value small-$x$. On the other hand Eq. (25) describes the $x$-evolutions of gluon distribution function at LO for a given value of $Q^2$.

\subsection{Analysis of gluon distribution function at NLO and NNLO:}

\ Substituting the splitting functions upto NLO [25-27] and NNLO [28, 29] in Eq. (1) and simplifying, we get the DGLAP equations for gluon distribution function at NLO and NNLO as
\begin{eqnarray}\frac{{\partial}G(x,t)}{{\partial}t}&=&\frac{\alpha_s(t)}{2\pi}\Big[6\Big(\frac{11}{12}-\frac{N_f}{18}+\ln(1-x)\Big)G(x,t)+6I_1^g(x,t)\Big]\nonumber\\
&& +\: \Big(\frac{\alpha_S(t)}{2\pi}\Big)^2I_2^g(x,t),\end{eqnarray}
\begin{eqnarray}\frac{{\partial}G(x,t)}{{\partial}t}&=&\frac{\alpha_s(t)}{2\pi}\Big[6\Big(\frac{11}{12}-\frac{N_f}{18}+\ln(1-x)\Big)G(x,t)+6I_1^g(x,t)\Big]\nonumber\\
&& +\: \Big(\frac{\alpha_S(t)}{2\pi}\Big)^2I_2^g(x,t)+\Big(\frac{\alpha_S(t)}{2\pi}\Big)^3I_3^g(x,t),\end{eqnarray}
where the integral functions $I_2^g(x,t)$ and $I_3^g(x,t)$ are defined as
\begin{equation}
I_2^g(x,t)=\int_x^1d\omega\Big[P_{gg}^{1}(\omega)G\big(\frac{x}{\omega},t\big)+A_{\omega}F_2^S\Big(\frac{x}{\omega},t\Big),\end{equation}
\begin{equation}{{I_3^g(x,t)}={\int_{x}^{1}{d{\omega}}\Big[P_{gg}^{2}(\omega)G\Big(\frac{x}{\omega},t\Big)]}}.\end{equation}
The explicit forms of $P_{gg}^{1}(\omega)$, $P_{gg}^{2}$ and $A_{\omega}$ are given in Appendix B. In the NNLO analysis we overlook the quark contribution to the gluon distribution function. The reason behind this approximation is that at very small values of $x$, the gluons, being the most abundant parton, dominate over the quarks. Moreover, it simplifies the calculations involving the NNLO splitting functions which otherwise are very complicated to solve analytically.

\ Following the same procedure as in LO, the Eqs.(26) and (27) can be simplified as
\begin{equation}{-\Big(\frac{t^2}{t-b\ln{t}}\Big)\frac{\partial{G(x,t)}}{\partial{t}}+L_2^g(x) \frac{\partial{G(x,t)}}{\partial{x}}+M_2^g(x)G(x,t)=0},\end{equation}
\begin{equation}{-\bigg(\frac{t^2}{t-b\ln{t}+b^2(\ln^2{t}-\ln{t}-1)+c}\bigg)+L_3^g(x) \frac{\partial{G(x,t)}}{\partial{x}}+M_3^g(x)G(x,t)=0}.\end{equation}
Here \begin{equation}L_2^g(x)=\frac{2}{\beta_0}\Big[6\Big(A_2^g(x)+K_1(x)A_4^g(x)\Big)+T_0\Big(B_2^g(x)+K_1(x)B_4^g(x)\Big)\Big],\end{equation}
\begin{eqnarray}
M_2^g(x)&=&\frac{2}{\beta_0}\Big[6\Big(A_1^g(x)+K_1(x)A_3^g(x)+\frac{\partial{K_1(x)}}{\partial{x}}A_4^g(x)\Big)\nonumber\\
&& +\: T_0 \Big(B_1^g(x)+K_1(x)B_3^g(x)+\frac{\partial{K_1(x)}}{\partial{x}}B_4^g(x)\Big)\Big],
\end{eqnarray}
\begin{eqnarray}L_3^g(x)&=&\frac{2}{\beta_0}\Big[6\Big(A_2^g(x)+K_1(x)A_4^g(x)\Big)+T_0\Big(B_2^g(x)+ K_1(x)B_4^g(x)\Big)\nonumber\\
&& +\: T_1C_2^g(x)\Big],
\end{eqnarray}
\begin{eqnarray}
M_3^g(x,t)&=&\frac{2}{\beta_0}\Big[6\Big(A_1^g(x)+K_1(x)A_3^g(x)+\frac{\partial{K_1(x)}}{\partial{x}}A_4^g(x)\Big)+T_0 \Big(B_1^g(x)\nonumber\\
&& +\: K_1(x)B_3^g(x)+\frac{\partial{K_1(x)}}{\partial{x}}B_4^g(x)\Big)+T_1C_1^g(x)\Big],
\end{eqnarray}
where $B_i^{g}(x)$, ($i$=1,2,3,4) and $C_i^{g}(x)$, ($i$=1,2) are functions of $x$ (see Appendix A).
To obtain an analytical solution of Eq.(4.29) we consider the numerical parameters $T_0$ and $T_1$ such that  ${T^2(t)=T_0T(t)}$ and ${T^3(t)=T_1T(t)}$, where $T(t)$={\large $\frac{\alpha_S(t)}{2\pi}$}. The value of the parameters $T_0$ and $T_1$ are determined by phenomenological analysis from a particular range of $Q^2$ under study and by an appropriate choice of $T_1$ as well as $T_1$ the errors can be reduced to a minimum. 

\ Thus proceeding in the same way we solve Eq.(30) to obtain the $t$ or $Q^2$ and $x$-evolutions of gluon distribution function at NLO as
\begin{equation} {G(x,t)=G(x,t_0)\Big(\frac{t^{1+{b/t}}}{t_0^{1+{b/t_0}}}\Big)\cdot{\exp}\Big(\frac{b}{t}-\frac{b}{t_0}\Big)}\end{equation}
and
\begin{equation} {G(x,t)=G(x_0,t)\cdot{\exp}\Big[\int_{x_0}^{x}\big(\frac{1}{{L_2^g}(x)}- \frac{{M_2^g}(x)}{{L_2^g}(x)}\big)dx\Big]}\end{equation}
with $b=\frac{\normalsize\beta_1}{\normalsize\beta_0^2}$. The input functions $G(x,t_0$) and $G(x_0,t)$ can be determined by applying the initial conditions at $t=t_0$ as well as at $x=x_0$ as in the previous case.

Similarly solving Eq.(31) the $t$ or $Q^2$ and $x$-evolutions of gluon distribution function at NNLO can be evaluated as
\begin{equation}G(x,t)=G(x,t_0)\Big(\frac{t^{1+{(b-b^2)/t}}}{t_0^{1+{(b-b^2)/t_0}}}\Big)\cdot{\exp}\Big(\frac{b-c-b^2\ln^2{t}}{t}-\frac{b-c-b^2\ln^2{t_0}}{t_0}\Big)
\end{equation}
and
\begin{equation} {G(x,t)=G(x_0,t)\cdot{\exp}\Big[\int_{x_0}^{x}\big(\frac{1}{{L_3^g}(x)}-\frac{{M_3^g}(x)}{{L_3^g}(x)}\big)dx\Big]},\end{equation}
where $b=${\large$\frac{\beta_1}{\beta_0^2}$} and $c=${\large$\frac{\beta_2}{\beta_0^3}$}. The input functions $G(x,t_0$) and $G(x_0,t)$ can be determined by applying the initial conditions at $t=t_0$ as well as at $x=x_0$ as earlier.

\section{Result and discussion}

\ In this paper, we obtain the $Q^2$ or $t$ ($t=\ln(Q^2/\Lambda^2)$) and $x$-evolutions of the gluon distribution function solving the DGLAP evolution equation for gluon density up to NNLO approximation. 
The analysis is performed in the range $10^{-4}\leq{x}\leq{0.1}$ and $5\leq{Q^2}\leq{110}$ GeV$^2$.
The computed results of gluon distribution function at LO, NLO and  NNLO are compared with the available GRV1998NLO [15], MRST2004NNLO [16], MSTW2008NNLO [17] and JR09NNLO [18] global QCD analysis. We also compare our results with the results of the BDM model [14].
THE GRV1998 global analysis is based on H1 and ZEUS high precision data [31, 32] on $G(x,Q^2)$ which radiatively generates the LO and NLO dynamical parton densities from valence-like inputs. JR09NNLO utilize the recent DIS measurements and data on hadronic dilepton production and determine, at 3-loop, the dynamical parton distributions of the nucleon generated radiatively from valence-like positive input distributions at an optimally chosen low resolution scale $Q_0^2<1$ GeV$^2$. The MRST2004NNLO provides a more physical parametrization of the gluon distribution at NLO and NNLO for global parton analyses of DIS and related hard scattering data. Here the gluon distribution at large-$x$ in the $\overline{MS}$ scheme is driven by the valence quarks producing a better description of the Tevatron inclusive jet data [33-35]. MSTW2008NNLO is an updated global analysis of hard-scattering data [33-41] within the standard framework of leading-twist fixed-order collinear factorisation in the $\overline{MS}$ scheme and presents the LO, NLO and NNLO parton distribution functions. These sets are a major update to all the previously available MRST sets. The BDM model obtains an  analytic solution for the LO  gluon distribution function directly from the proton structure function using the accurate Froissart-bound [42] type parametrization of proton  structure function. In this model, it is shown that using an analytic expression that successfully reproduces the known experimental data for proton structure function in a domain $x_{min}(Q^2)\leq{x}\leq{x_{max}}(Q^2)$ and $Q^2_{min}\leq{Q^2}\leq{Q^2_{max}}$ in DIS, the gluon distribution $G(x,Q^2)$ can be uniquely determined in the same domain of $x$ and $Q^2$.
In all the graphs, the lowest-$Q^2$ and highest-$x$ points are taken as input for $G(x,t_0)$ and $G(x_0,t)$ respectively. 
We consider a function $K_1(x)$ which relates the gluon distribution and the singlet structure function. For simplicity we consider the function $K_1(x)=K_1$, where $K_1$ is a constant parameter. The acceptable range of the arbitrary constant $K_1$ is found to be $0.14\leq{K_1}\leq{0.85}$. In each figure the dot lines represent our LO results, the dash-dot lines represent our NLO results whereas the solid lines represent the NNLO results. As expected the improvement is found to be better at NNLO than at NLO and LO.
\begin{figure}[!htb]
\centering
\includegraphics[width=4.5in,height=3.5in]{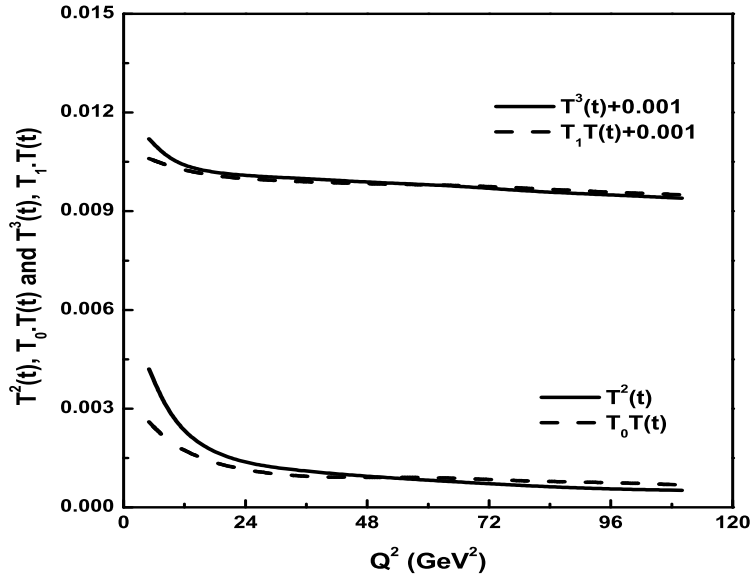}
\caption{\footnotesize{Comparison of $T^2$ and $T_0.T(t)$ as well as $T^3$ and $T_1.T(t)$ versus $Q^2$.}}
\label{fig:1}
\end{figure}

\ In the calculation of gluon distribution function at NLO and NNLO, we consider two numerical parameters $T_0$ and $T_1$ to linearise the equations in $\alpha_s$. 
These numerical parameters are obtained for a particular range of $Q^2$ under study. Figure 1 shows the plot of $T^2(t)$ and $T_0T(t)$ as well as $T^3(t)$ and $T_1T(t)$ versus $Q^2$ in the range $2<Q^2<110$ GeV$^2$. It is observed that for $T_0=0.035$ and $T_1=0.0042$ the differences between $T^2(t)$ and $T_0T(t)$ as well as $T^3(t)$ and $T_1T(t)$ becomes negligible in the $Q^2$ range under study. Therefore, the consideration of the parameters $T_0$ and $T_1$ does not induce any unexpected change in our results.
\begin{figure}[!htb]
\centering
\includegraphics[width=2.6in, height=2.7in]{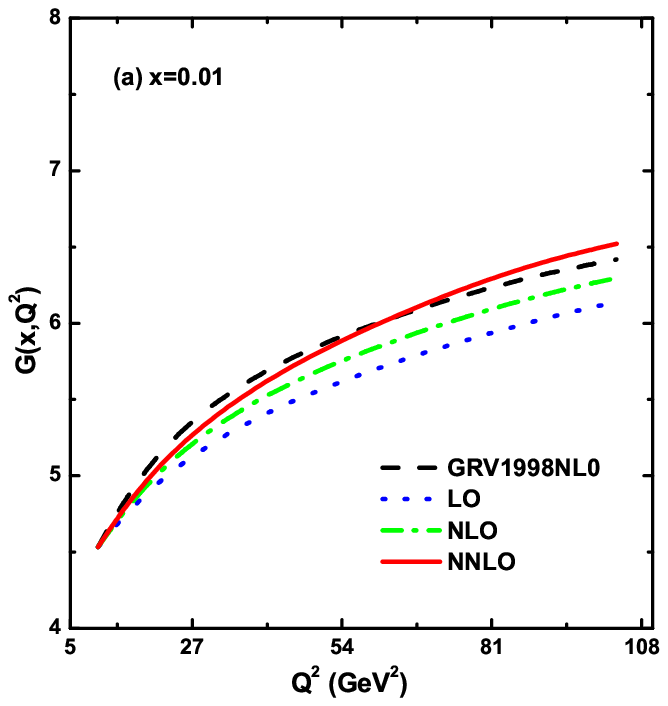}
\includegraphics[width=2.6in, height=2.7in]{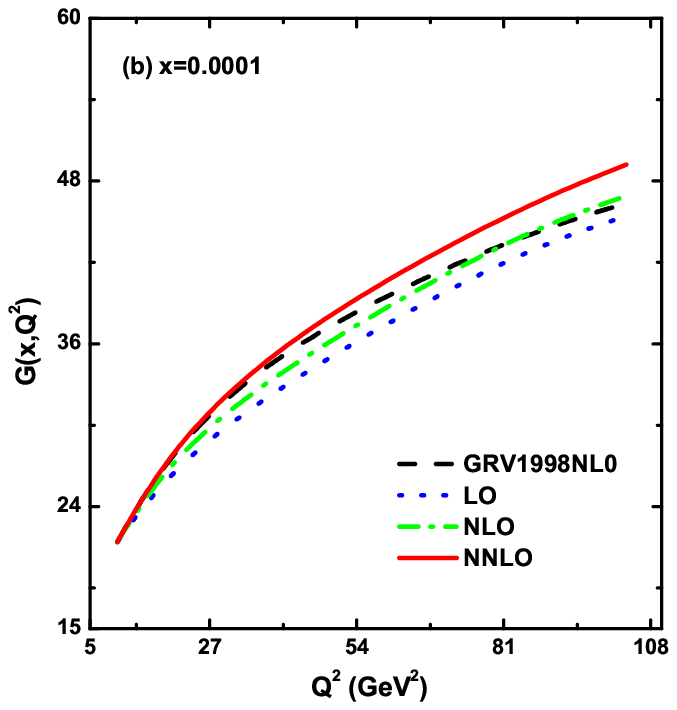}
\caption{\footnotesize {${Q^2}$ evolution of gluon distribution functions at LO, NLO and NNLO compared with GRV1998NLO for two fixed values $x$. The dot lines represent the LO results (Eq.(24)), dash-dot lines represent the NLO results (Eq.(36)) and solid lines represent the NNLO results (Eq.(38)).}}
\label{fig:1}
\end{figure}

\begin{figure}[!htb]
\centering
\includegraphics[width=2.6in, height=2.7in]{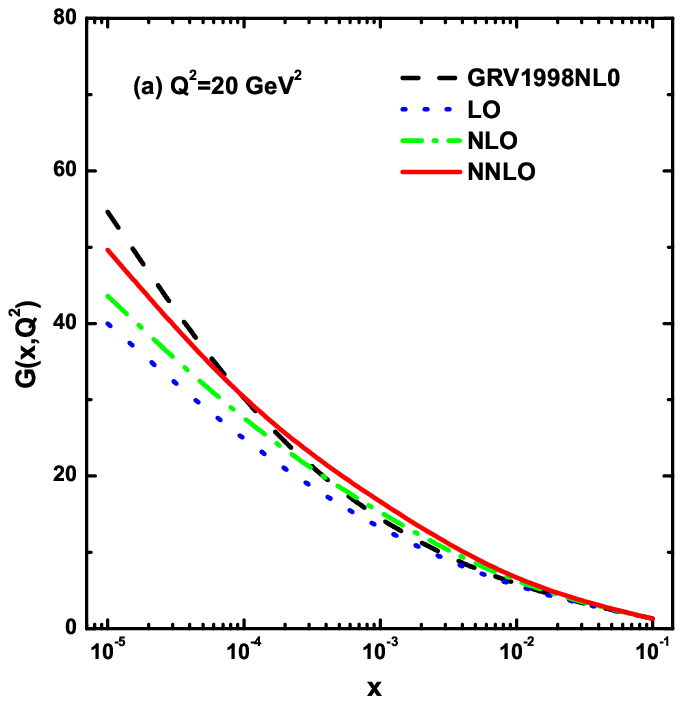}
\includegraphics[width=2.6in, height=2.7in]{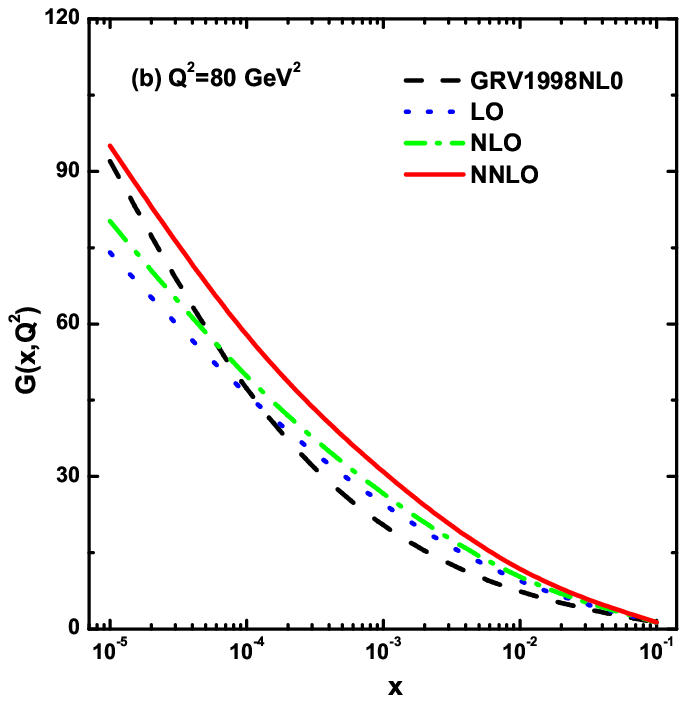}
\caption{\footnotesize{$x$ evolution of gluon distribution functions at LO, NLO and NNLO compared with GRV1998NLO data for two fixed $Q^2$. The dot lines represent the LO results (Eq.(25)), dash-dot lines represent the NLO results (Eq.(37)) and solid lines represent the NNLO results (Eq.(39)).}}
\label{fig:1}
\end{figure}

\begin{figure}[!htb]
\centering
\includegraphics[width=2.6in, height=2.7in]{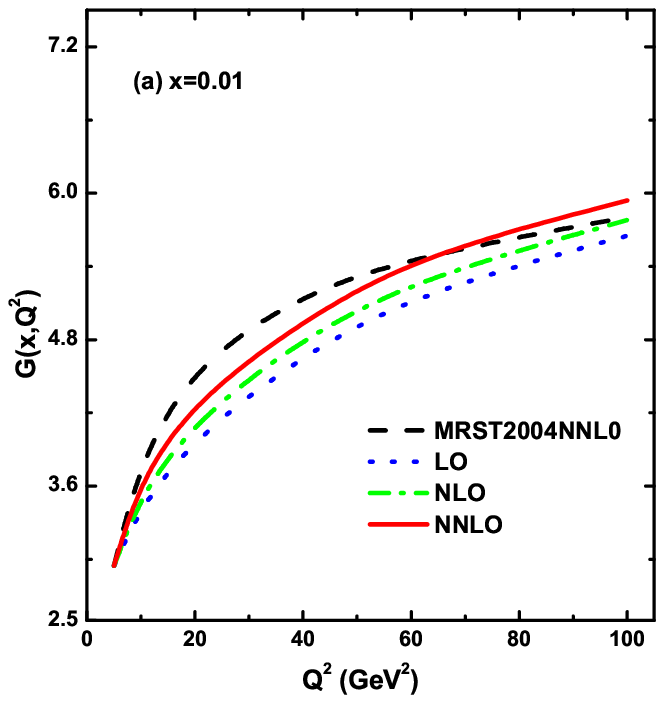}
\includegraphics[width=2.6in, height=2.7in]{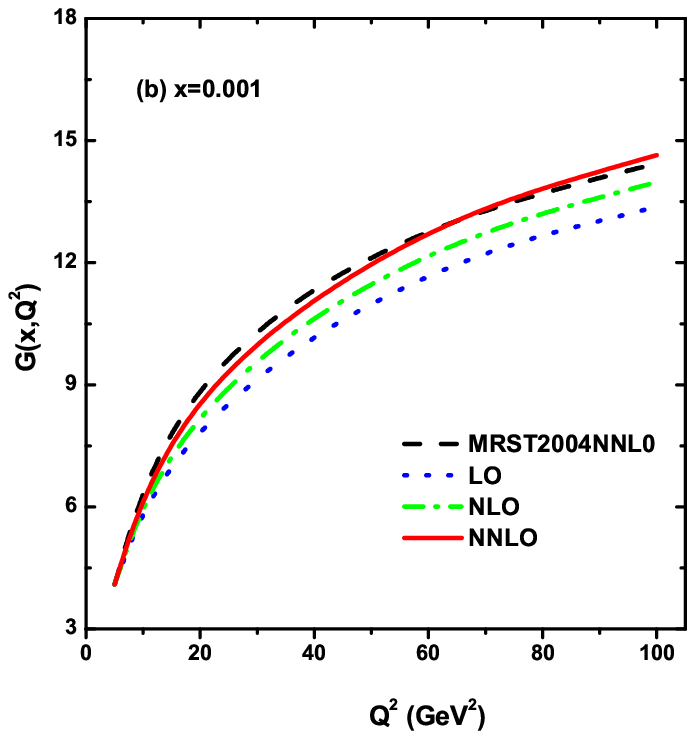}
\includegraphics[width=2.6in, height=2.7in]{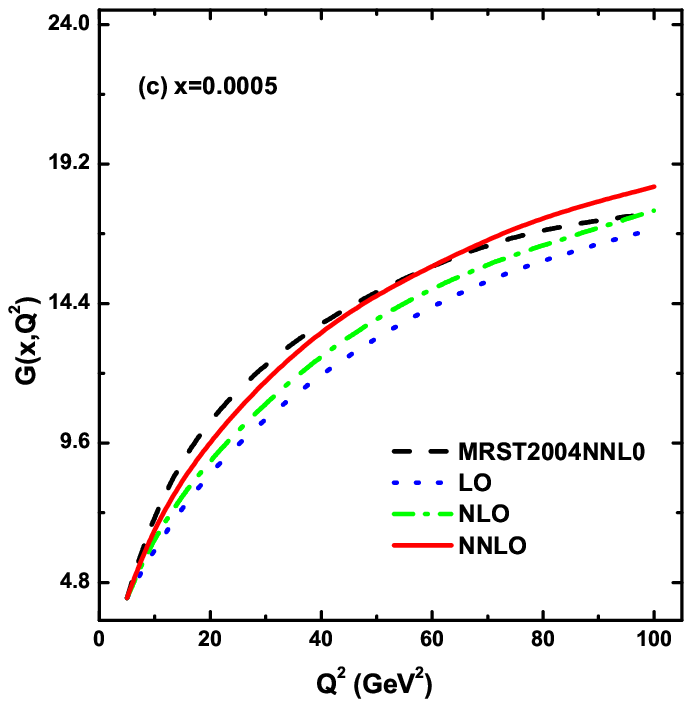}
\includegraphics[width=2.6in, height=2.7in]{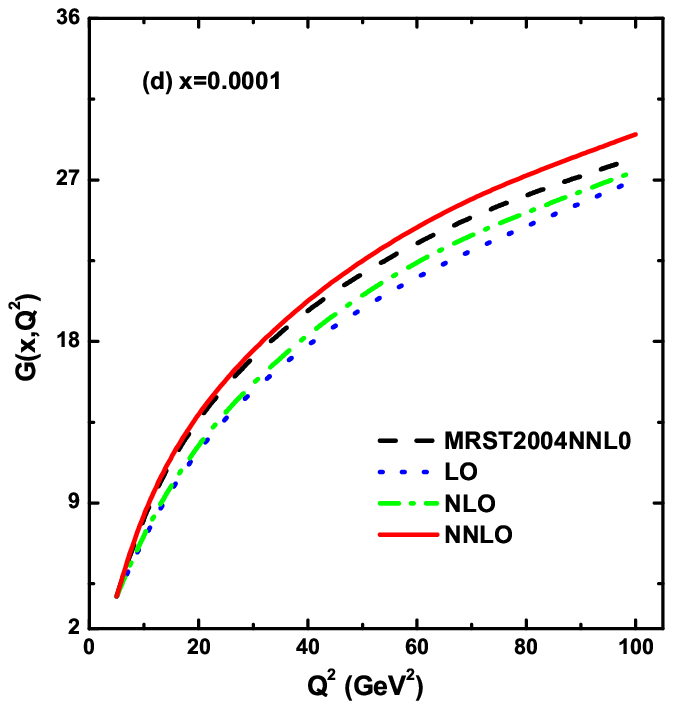}
\caption{\footnotesize{Comparison of ${Q^2}$ evolution  of gluon distribution functions at LO, NLO and NNLO with MRST2004NNLO parametrization for four fixed values $x$, viz. The dot lines represent the LO results (Eq.(24)), dash-dot lines represent the NLO results (Eq.(36)) and solid lines represent the NNLO results(Eq.(38)).}}
\label{fig:1}
\end{figure}

\ Figures 2, 4 and 6 represent the comparison of our computed results of $Q^2$ evolution of gluon distribution function $G(x,Q^2)$ calculated from Eqs.(24), (36) and (38) at LO, NLO and NNLO with the GRV1998NLO, MRST2004NNLO and MSTW2008NNLO global parametrizations respectively. On the other hand,  Figures 3, 5 and 7 depict the $x$-evolutions of $G(x,Q^2)$ at LO, NLO and NNLO obtained from Eqs. (25), (37) and (39) respectively compared with the GRV1998NLO, MRST2004NNLO and MSTW2008NNLO parametrizations respectively. It is very interesting to find that the NNLO approximation improves the agreement of our predicted values of $G(x,Q^2)$ with the results of different global PDF groups.
\begin{figure}[!htb]
\centering
\includegraphics[width=2.6in, height=2.7in]{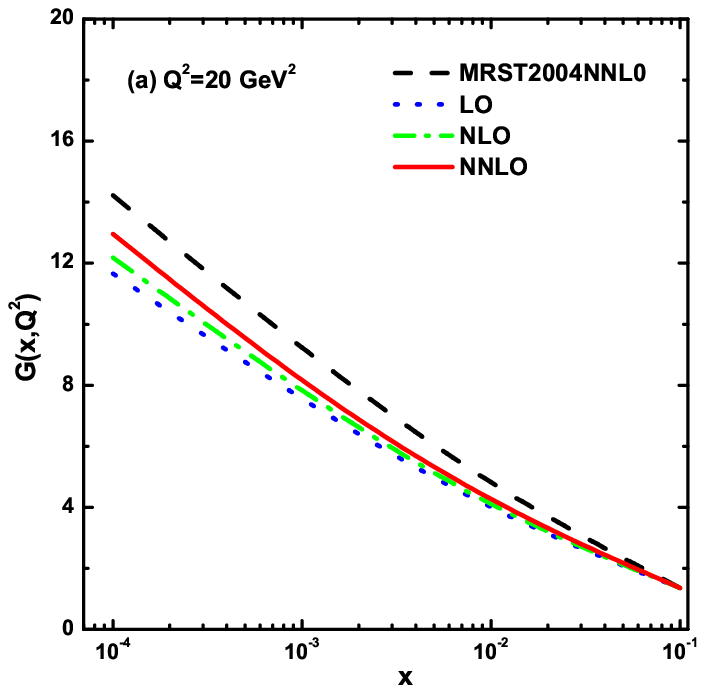}
\includegraphics[width=2.6in, height=2.7in]{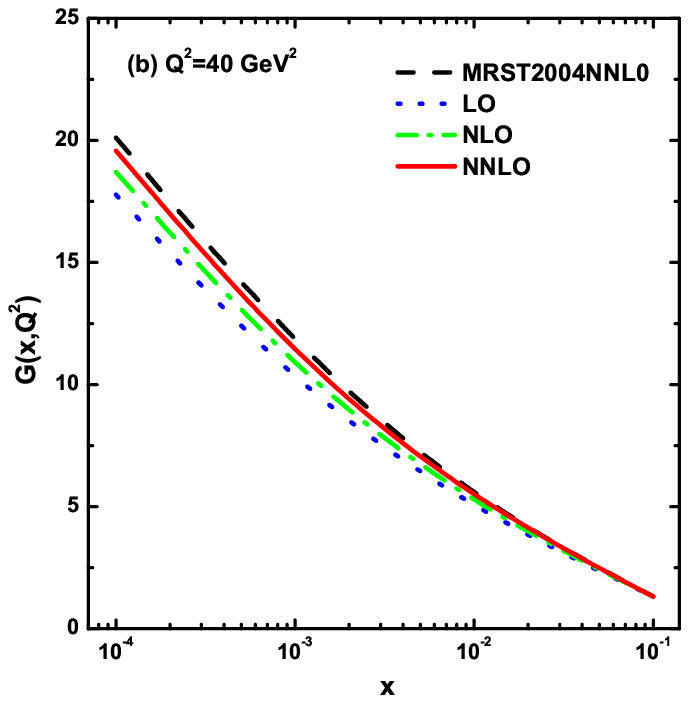}
\includegraphics[width=2.6in, height=2.7in]{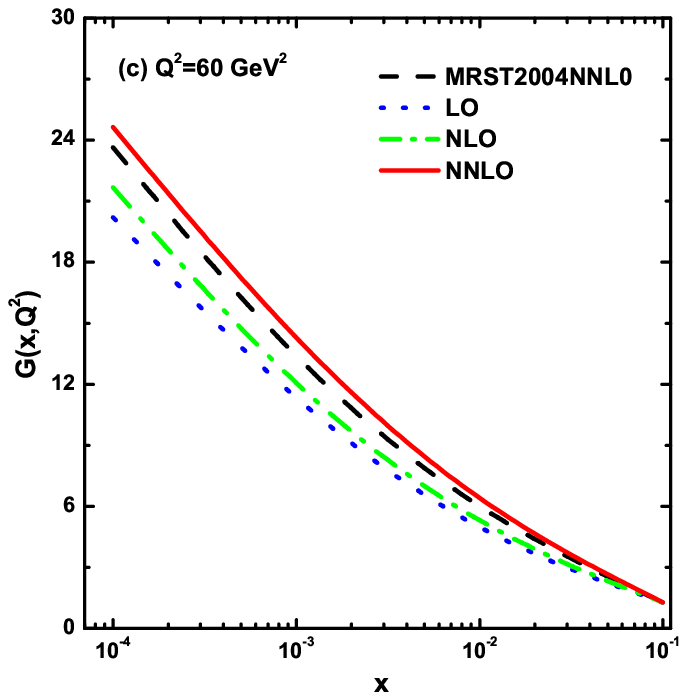}
\includegraphics[width=2.6in, height=2.7in]{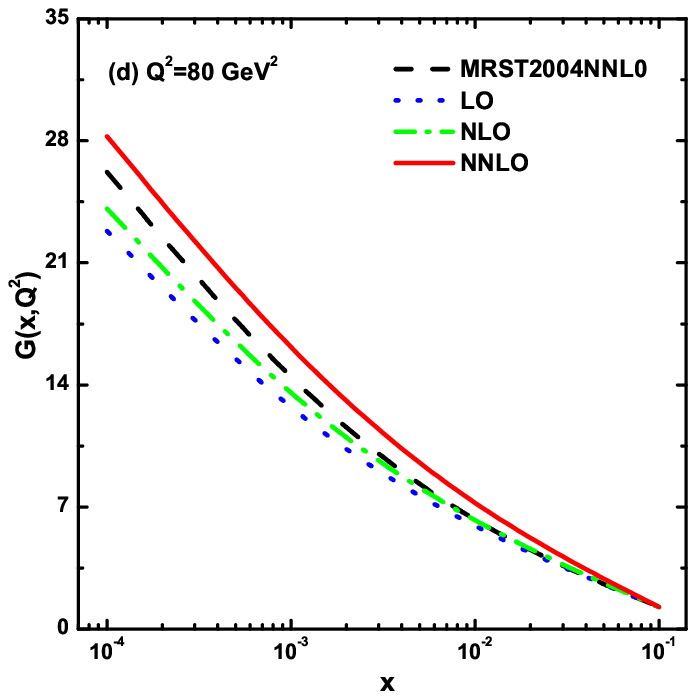}
\caption{\footnotesize{Comparison of the values of gluon distribution function at LO, NLO and NNLO plotted against $x$ with the MRST2004NNLO parametrization for four representative $Q^2$. The dot lines represent the LO results (Eq.(25)), dash-dot lines represent the NLO results (Eq.(37)) and solid lines represent the NNLO results (Eq.(39)).}} 
\label{fig:1}
\end{figure}

\begin{figure}[!htb]
\centering
\includegraphics[width=2.6in, height=2.7in]{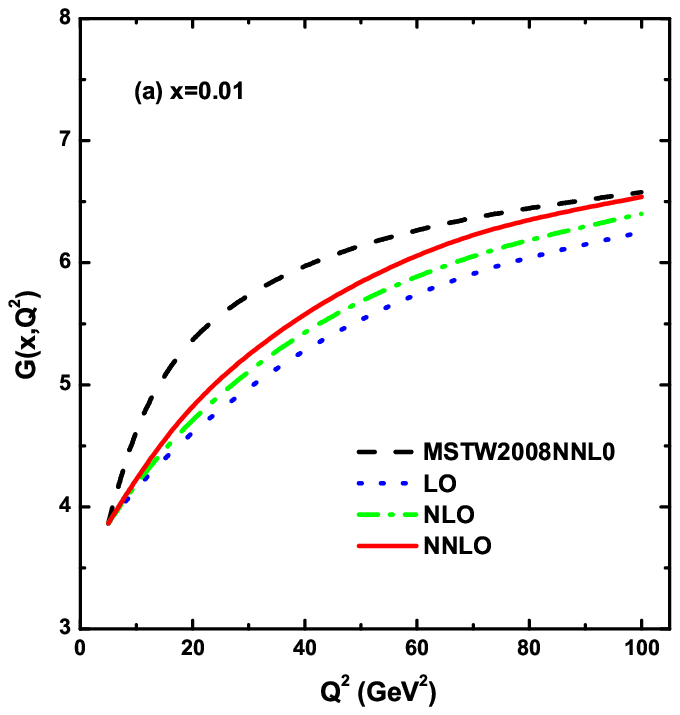}
\includegraphics[width=2.6in, height=2.7in]{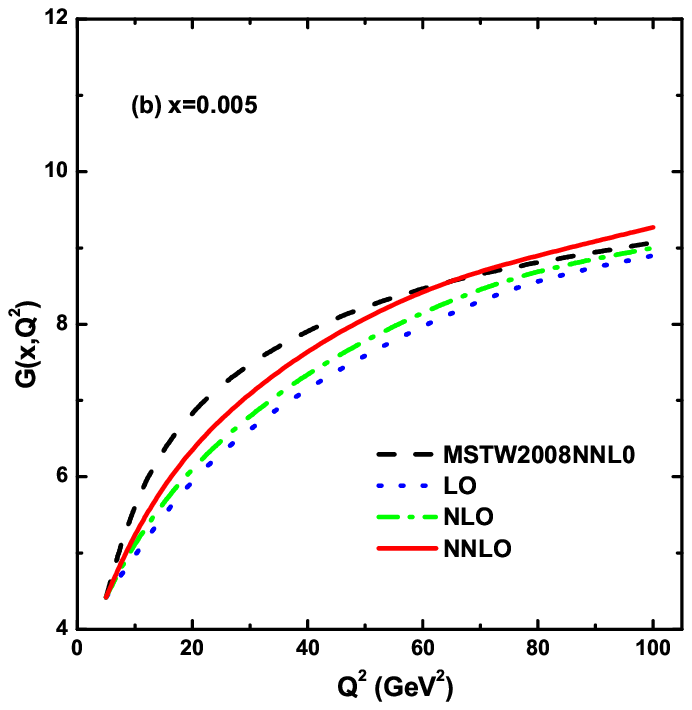}
\includegraphics[width=2.6in, height=2.7in]{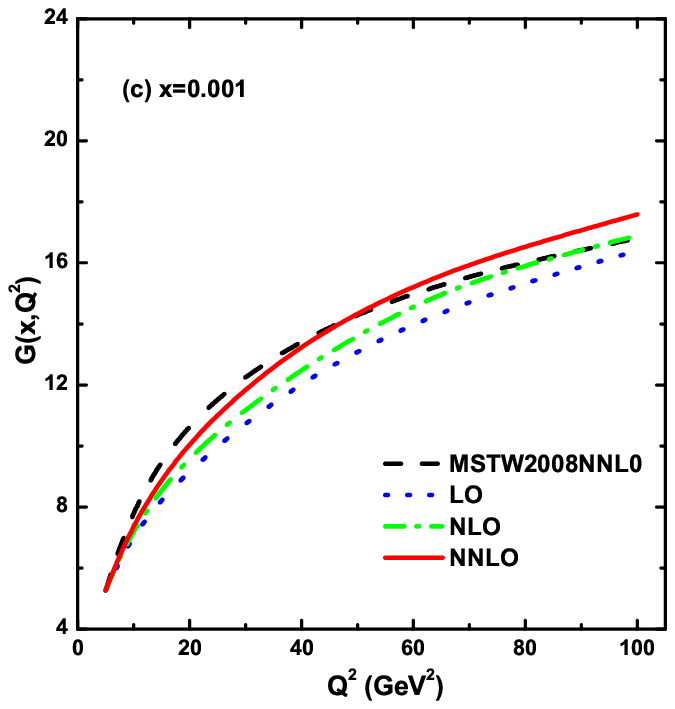}
\includegraphics[width=2.6in, height=2.7in]{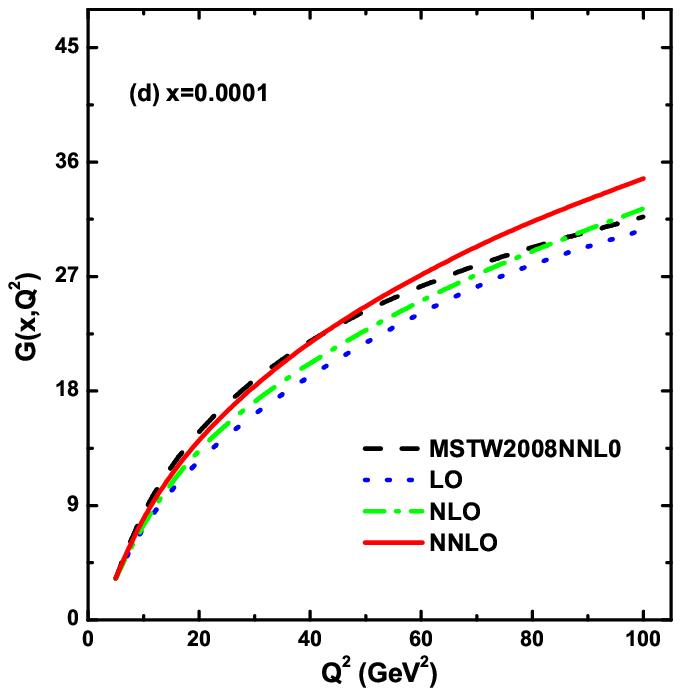}
\caption{\footnotesize{$Q^2$ evolution of gluon distribution function at LO, NLO and NNLO compared with MSTW2008NNLO parametrization for four fixed $x$ values. The dot lines represent the LO results (Eq.(24)), dash-dot lines represent the NLO results (Eq.(36)) and solid lines represent the NNLO results (Eq.(38)).}}
\label{fig:1}
\end{figure}

\begin{figure}[!htb]
\centering
\includegraphics[width=2.6in, height=2.7in]{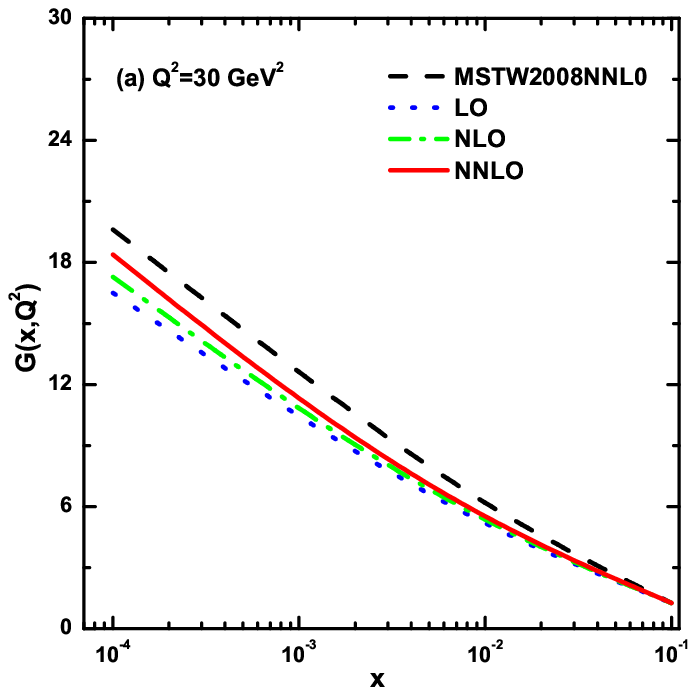}
\includegraphics[width=2.6in, height=2.7in]{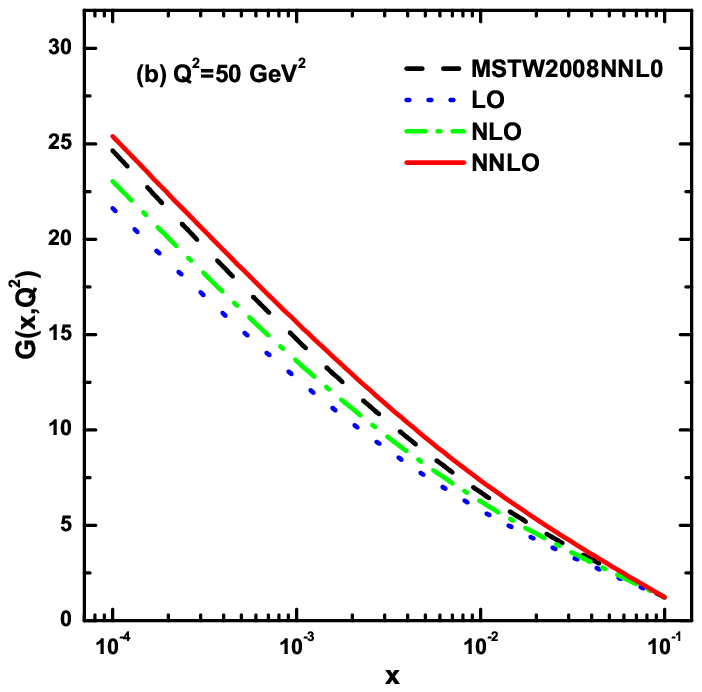}
\includegraphics[width=2.6in, height=2.7in]{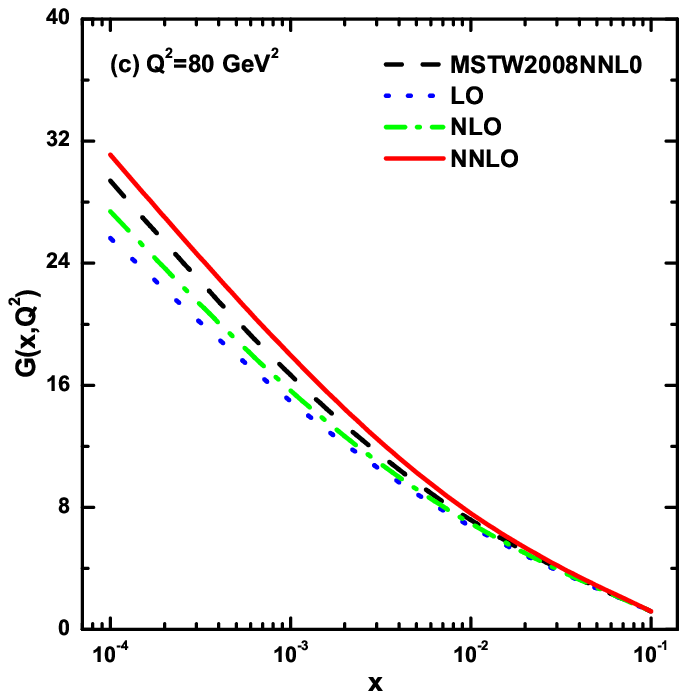}
\includegraphics[width=2.6in, height=2.7in]{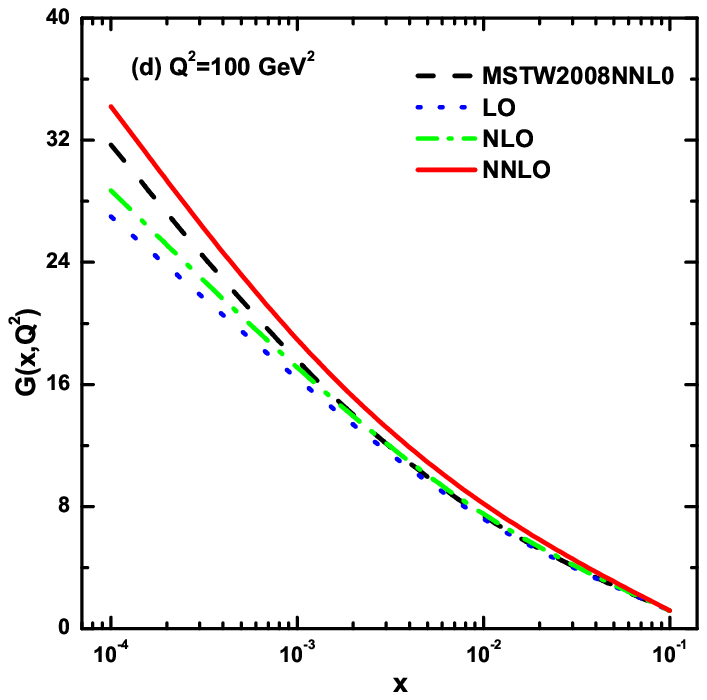}
\caption{\footnotesize{Comparison of the values of gluon distribution function at LO, NLO and NNLO plotted against $x$ with the MRST2004NNLO parametrization for four representative $Q^2$. The dot lines represent the LO results (Eq.(25)), dash-dot lines represent the NLO results (Eq.(37)) and solid lines represent the NNLO results (Eq.(39)).}}
\label{fig:1}
\end{figure}

\begin{figure}[!htb]
\centering
\includegraphics[width=2.6in, height=2.7in]{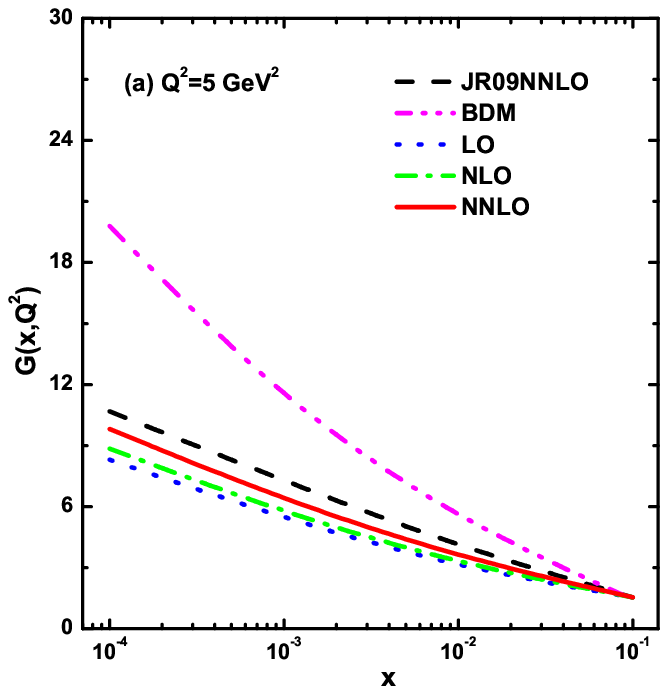}
\includegraphics[width=2.6in, height=2.7in]{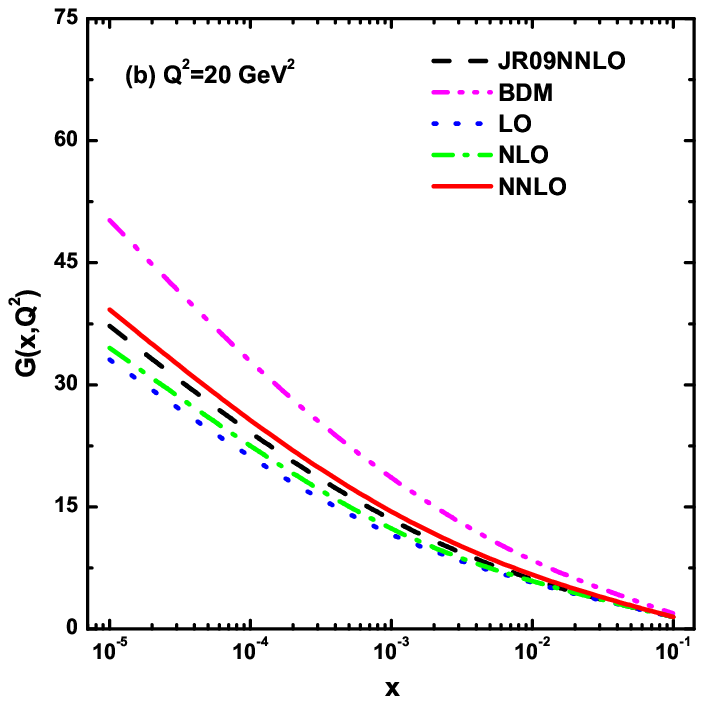}
\caption{\footnotesize{Comparison of our predictions of gluon distribution function at LO, NLO and NNLO with JR09NNLO global fit as well as with the BDM model for two representative $Q^2$. The dot lines are the LO results of obtained from (Eq.(25)), dash-dot lines are NLO results from (Eq.(37)) and solid lines are the NNLO results from (Eq.(39)). The dash curves are from the JR09NNLO parametrization and dash-dot-dot curves are the results of BDM model.}}
\label{fig:1}
\end{figure}

\ We also compare the predicted results of $x$ dependence of gluon distribution function $G(x,Q^2)$ with the JR09NNLO global parton analysis [18] as well as with the results of BDM model [14]. This comparison is portrayed in Figure 8 where we plot the computed values of $G(x,Q^2)$ at LO, NLO and NNLO using Eqs. (25), (37) and (39) versus $x$ in the range ${10^{-4}\leq{x}\leq{0.1}}$ for $Q^2=5$ GeV$^2$ and $Q^2=20$ GeV$^2$. Our predictions of $G(x,Q^2)$ at NNLO show very good agreement with the JR09NNLO. Our results also show similar behaviour with those of BDM model, however the BDM model gives much larger gluon distribution towards small $x$. Figures indicate that the compatibility of our predictions with the JR09NNLO parametrization much better that that of BDM model.

\begin{table}[hth]
\caption{$\chi^2$ text for $G(x,Q^2)$}   
\centering                         
\vspace{4pt}
\begin{tabular}{ |c|c|c|c|c| }        
\hline
\bf{Order}  & \bf{GRV1998}  &   \bf{MRST2004} &  \bf{MSTW2008} & \hspace{10pt} \bf{JR09} \hspace{10pt} \\ [1.5ex]    
\hline                              
LO & 4.03 & 2.34 & 3.18 & 2.38 \\[0.5ex]          
\hline
NLO & 3.16 & 1.63 & 2.16 & 1.24 \\ [0.5ex]         
\hline                              
NNLO & 2.96 & 0.85 & 1.72 & 0.61 \\ [0.5ex]
\hline
\end{tabular}
\end{table}

\ To check the compatibility of our results of gluon distribution function at LO, NLO and NNLO respectively with different parametrizations, we perform a $\chi^2$ test shown in Table 1. From this we observe that our results are almost comparable with different parametrizations and the inclusion of NNLO contributions improve the consistency.

\section{Summary}

\ To summarise the evolution of gluon distribution function with respect to $x$ and $Q^2$ at LO, NLO and NNLO are presented by solving the DGLAP evolution equation for gluon distribution analytically. Here the DGLAP equation is first transformed into a partial differential equation in the two variables $x$ and $Q^2$ by using the Taylor series expansion valid at small-$x$. Following this the resulting equation is solved at LO, NLO and NNLO respectively by the Lagrange's auxiliary method to obtain the $Q^2$ and $x$ evolutions of the gluon distribution function.
We compare our predictions with the GRV1998NLO, MRST2004NNLO, MSTW2008NNLO and JR09NNLO global QCD analysis as well as with the BDM model. The obtained results of $G(x,Q^2)$ can be described within the framework of perturbative QCD. Our results show that at fixed $x$ the gluon distribution function increases with increasing $Q^2$ whereas at fixed $Q^2$ it decreases as $x$ decreases which is in agreement with perturbative QCD fits at small-$x$. We perform our analysis in the $x$ and $Q^2$ range, viz. $10^{-4}\leq{x}\leq{0.1}$ and $5\leq{Q^2}\leq{110}$ GeV$^2$ and note that in this domain our predicted solutions are comparable with different global analysis of parton distributions. We consider the function $K_1(x)=K_1$, where $K_1$ is a constant parameter, in defining the relation between gluon and singlet structure functions and obtain our best fitted results in the range $0.14\leq{K_1}\leq{0.85}$. Moreover we consider the numerical parameters $T_0$ and $T_1$ to linearise the equations at NLO and NNLO in $\alpha_s$. These parameters are chosen from phenomenological analysis for a particular range of $Q^2$ under study and therefore, the use of the parameters $T_0$ and $T_1$ does not produce any abrupt change in our results.
From our phenomenological analysis we observe that our computed results of gluon distribution function at NNLO show significantly better agreement with different parameterizations than those of LO and NLO. Thus we can say that the NNLO approximation has appreciable contribution to the gluon distribution function in the particular range of $x$ and $Q^2$ under study. However, in the very small-$x$ region, where the number density of gluons become very high, the gluon recombination processes may take place inducing nonlinear corrections to the QCD evolution and in that case the nonlinear GLR-MQ evolution equation may provide a good description of the high density QCD at very small-$x$, which is discussed elsewhere [52-55].

{\bf{\large{Appendix:A}}}

The explicit forms of the functions  $A_i^g(x)$, $B_i^g(x)$ ($i$=1,2,3,4) and $C_i^g(x)$ ($i$=1,2)  are
\begin{equation}
A_1^g(x)=-\frac{11}{6}+2x-\frac{1}{2}x^2+\frac{1}{3}x^3-\ln(x),
\end{equation}
\begin{equation}
A_2^g(x)=1+\frac{4}{3}x-3x^2+x^3-\frac{1}{4}x^4+2x\ln(x).
\end{equation}
\begin{equation}
A_3^g(x)=\frac{2}{9}(-\frac{3}{2}+2x-\frac{1}{2}x^2-2\ln(x)),
\end{equation}
\begin{equation}
A_4^g(x)=\frac{2}{9}(2+\frac{1}{2}x-3x^2+\frac{1}{2}x^3+4x\ln(x),
\end{equation}
\begin{equation}
B_1^g(x)=-\frac{52}{3}\ln(x),
\end{equation}
\begin{equation}
B_2^g(x)=-\frac{52}{3}(1-x+x\ln(x)),
\end{equation}
\begin{equation}
B_3^g(x)=\int_{x}^{1}A(\omega)d\omega,
\end{equation}
\begin{equation}
B_4^g(x)=x\int_{x}^{1}\frac{1-\omega}{\omega}A(\omega)d\omega,
\end{equation}
\begin{equation}C_1^g(x)=\int_x^1P_{gg}^2(\omega)d{\omega},\end{equation}
\begin{equation}C_2^g(x)=x\int_x^1\frac{1-\omega}{\omega}P_{gg}^2(\omega).\end{equation}\\

{\bf{\large{Appendix:B}}}

The functions involved in the DGLAP equations for gluon distribution functions at NLO are
\begin{eqnarray}
P_{gg}^1&=&C_FT_F(-16+8z+\frac{20}{3}z^2+\frac{4}{3}z-(6+10z)\ln(z)-(2+2z)lnz^2)\nonumber\\
&& +C_AT_F(2-2z+\frac{26}{9}(z^2-1/z)-\frac{4}{3}(1+z)\ln(z)-\frac{20}{9}P_{gg}(z))\nonumber\\
&& +C_A^2(\frac{27}{2}(1-z)+\frac{26}{9}(z^2-1/z)-(\frac{25}{3}-\frac{11}{3}z+\frac{44}{3}z^2)\ln(z)\nonumber\\
&& +4(1+z)\ln(z^2)+2P_{gg}(-z)S_2(z)+(\frac{67}{9}-4\ln(z)\ln(1-z)\nonumber\\
&& +\ln(z^2)-\frac{\pi^{2}}{3})P_{gg}(z)).
\end{eqnarray}
\begin{equation} A(\omega)=C_F^2A_1(\omega)+C_FC_GA_2(\omega)+C_FT_RN_FA_3(\omega)
\end{equation}
where
\begin{eqnarray} A_1(\omega)&=&-\frac{5}{2}-\frac{7}{2}\omega+(2+\frac{7}{2}\omega)+
(-1+\frac{\omega}{2})\ln^2\omega-2\omega.\ln(1-\omega)\nonumber\\
&& +(-3\ln(1-\omega)-\ln^2(1-\omega))\frac{1+(1-\omega)^2}{\omega},
\end{eqnarray}
\begin{eqnarray} A_2(\omega)&=&\frac{28}{9}+\frac{65}{18}.\omega+\frac{44}{9}\omega^2+(-12-5\omega-\frac{8}{3}\omega^2)\ln\omega+(4+\omega)\ln^2\omega\nonumber\\
&& +2\omega\ln(1-\omega) +(-2\ln\omega\ln(1-\omega)+\frac{1}{2}\ln^2\omega+\frac{11}{3}\ln(1-\omega)\nonumber\\
&& +\ln^2(1-\omega)-\frac{1}{6}\pi^2+\frac{1}{2})\frac{1+(1-\omega)^2}{\omega}\nonumber\\
&& -\frac{1+(1+\omega)^2}{\omega}\int_{\omega/1+\omega}^{1/1+\omega}\frac{dz}{z}\ln(\frac{1-z}{z}),
\end{eqnarray}
\begin{equation} A_3(\omega)=-\frac{4}{3}\omega-(\frac{20}{9}+\frac{4}{3}\ln(1-\omega))(\frac{1+(1-\omega)^2}{\omega}).
\end{equation}\\

The functions involved in the DGLAP equations for gluon distribution functions at NNLO are
\begin{eqnarray}
P_{gg}^2&=&2643.524D_0+4425.894\delta(1-z)+3589L_1-20852+3968z-3363z^2\nonumber\\
&& +4848z^3+L_0L_1(7305+8757L_0)+274.4L_0-7471L_0^2+72L_0^3-144L_0^4+\nonumber\\
&& \frac{142141}{z}+\frac{2675.81}{z}L_0+N_f(412.142D_0-528.723\delta(1-z)-320l_1\nonumber\\
&& -350.2+755.7z-713.8z^2+559.3z^3+L_0L_1(26.85-808.7L_0)+1541L_0\nonumber\\
&& +491.3L_0^2+\frac{832}{9}L_0^3+\frac{512}{27}L_0^4+\frac{182.961}{z}+\frac{157.271}{z}L_0)\nonumber\\
&& +N_f^2(-\frac{16}{9}D_0+6.4630\delta(1-z)-13.878+153.4z-187.7z^2+52.75z^3\nonumber\\
&& L_0L_1(115.6-85.25z+63.23L_0)-3.422L_0+9.680L_0^2-\frac{32}{27}L_0^3\nonumber\\
&& -\frac{680}{2431z)}
\end{eqnarray}
where, $L_0$=$\ln(z)$, $L_1$=$\ln(1-z)$ and $D_0$=$\frac{1}{(1-z)}$.

\end{document}